\documentstyle[12pt,preprint,tighten,floats,aps,epsf,psfig]{revtex}

\tighten
\begin{document}
\draft
\title{One-Loop Effects in Supergravity Models with an Additional U(1)}  
\author{D.A. Demir, N. K. Pak}
\address{Middle East Technical University, Department of Physics,
06531, Ankara, Turkey}
\date{\today} 
\maketitle
\begin{abstract}
For an Abelian extended Supergravity model, we investigate some 
important low energy parameters: $\tan\beta$, $Z-Z'$ mixing angle, 
lightest CP-even Higgs mass bound,  $Z'$ mass, and effective $\mu$ parameter.
By integrating the RGE's from string scale down to the weak scale we 
constuct the scalar potential, and analyze the quantities above at the 
tree- and one-loop levels by including the contributions of top squarks  
and top quark in the effective potential.\\
PACS: 04.65.+e, 12.60.Jv
\end{abstract} 
\newpage
\section{Introduction}
There are several reasons for considering additional $U(1)$ symmetries 
and their associated extra $Z$ bosons. Such additional $U(1)$'s arise after 
the breaking of GUT's (for example $E(6)$-based rank-5 models), or in 
string compactifications. In addition to justifying the underlying 
model, more importantly, additional $U(1)$'s would also solve the MSSM $\mu$ 
problem when broken around the weak scale. Indeed, as was already argued in 
\cite{cvetic-lang}, in a large class of string models, breaking scale 
of the extra $U(1)$'s come out to be below a $TeV$.

The phenomenologically viable models should satisfy two conditions 
at the string scale: Firstly, the extra $U(1)$ should be non-anomalous 
and should not acquire a mass from the string or hidden sector dynamics; 
namely, its mass must come from the gauge symmetry breaking in the 
observable sector. Secondly, all scalar soft mass-squareds must be 
positive and of similar magnitude. The latter holds in gravity-mediated 
SUSY breaking scheme, where the mass scale is given by the gravitino mass, 
not necessarily, so however, in gauge-mediated SUSY breaking schemes. 

Soft terms, parametrizing our ignorance of the origin of the SUSY 
breaking, can be obtained from a general supergravity (SUGRA) Lagrangian 
in $M_{Pl}\rightarrow \infty$ limit \cite{masiero,nilles}. Although, the 
minimal SUGRA predicts universal soft terms, in general SUGRA theories (see 
\cite{SW} and references therein), and in superstring theories \cite{BIMS} 
it is possible to have non-universal soft terms. Thus, considering such 
explicit examples, one is free to consider non-universal boundary 
conditions \cite{MN}, without referring to the particular case of 
universality.
  
For testing such extra $U(1)$ models in near-future machines, the 
tree-level potential is clearly not sufficient; one has to take into account 
the radiative corrections to have a meaningful model at these energies. 
Among other methods \cite{hollik}, the effective potential approach 
proved to be an elegant and simple way of icorporating the radiative 
corrections to the scalar potential \cite{CQW,QUIROS}, which we will 
adopt in this work as well.

This work is organized as foollows: In Sec. 2 we shall first 
describe the model at the SUGRA scale. Using one loop RGE's we shall 
obtain all the low energy potential parameters as functions of their 
SUGRA scale values. After discussing the requirements on the low energy 
potential for phenomenological viablility, we determine the appropriate 
SUGRA scale parameter space. We do this with minimal amount of 
non-universality. That is, we allow for non-universality only between 
Higgs doublets and remaining scalars; in particular, we choose doublet 
soft mass-squareds to be equal and one order of magnitude smaller than the 
others.
 
In Sec. 3 we consider the issue of radiative corrections. Out of all 
fields which can contribute to the effective potential, we consider top 
quark and  stops contributions, and neglect the remaining fields. 
We assume that the log effects which are accounted for in solving the 
RGE's, are enough to take into account the effects of Higgs, neutralino, 
chargino, and vector boson loops, at least for calculating the 
low-lying mass spectrum \cite{EKW2-Ell}. 
  
In Sec. 4 we work out the one-loop potential numerically, and graph the 
tree- and one-loop results together to enable a comparative discussion of 
the effects of the radiative corrections.

In Sec. 5 we discuss the results of the work in the light of near-future 
accelerators, and MSSM and NMSSM predictions.

\section{Low-Energy Tree-Level Potential}
As is well known, the fundamental SUGRA scale $M=M_{Pl}/\sqrt{8\pi}$ is 
approximately one order of magnitude larger than the MSSM coupling 
constant unification level $M_{U}\approx 10^{16}\, GeV$ \cite{lang-uni}.
However, the threshold effects \cite{lang-polonsky,kaplunovsky} can close 
the gap, and thus, in the following we shall choose the MSSM unification 
scale $M_{U}$ as the starting point of the analysis at which the initial 
conditions of the potential parameters are specified. We reconsider a 
general, anomaly-free, Abelian extended SUSY model which was discussed 
in \cite{Biz} already. The model is specified by an Abelian extension of 
the MSSM gauge group: $G=SU(3)_{c}\times SU(2) \times U(1)_{Y} \times 
U(1)_{Y'}$ with the couplings $g_{3}$, $g_{2}$, $g_{Y}$, $g_{Y'}$, 
respectively. The particle content for one family is given by the 
left-handed chiral superfields: $\hat{L} \sim (1,2,-1/2, {Q'}_{L})$, 
$\hat{E}^{c} \sim (1, 1, 1, {Q'}_{E})$, $\hat{Q} \sim (3, 2, 1/6, 
{Q'}_{Q})$, $\hat{U}^{c} \sim (\bar{3}, 1, -2/3, {Q'}_{U})$, 
$\hat{D}^{c} \sim (\bar{3}, 1, 1/3, {Q'}_{D})$. The Higgs sector 
contains the $SU(2)$ doublets $\hat{H}_{1} \sim (1, 2, -1/2, {Q'}_{1})$ and
$\hat{H}_{2} \sim (1, 2, 1/2, {Q'}_{2})$, and the SM-singlet 
$\hat{S}\sim (1, 1, 0, {Q'}_{S})$. 
   
The superpotential for the model is given by  
\begin{eqnarray}
f=h^{0}_{s}\,\hat{S}\hat{H}_{1}.\hat{H}_{2} + 
h^{0}_{t}\,\hat{U}^{c}\hat{Q}.\hat{H}_{2} 
\end{eqnarray}
where the family mixings are ignored. We kept only the top and Higgs 
Yukawa couplings, as the Yukawa coupings of the other fermions are much 
smaller.The existence of $U(1)_{Y'}$ group makes the model totally different 
from NMSSM by forbidding an elementary $\mu$ term and an $S^{3}$ term 
(so that the superpotential does not have a $Z_{3}$ symmetry).

Without specializing to a particular class of string compactifications, we 
parameterize the supersymmetry breaking by considering the most general 
soft supersymmetry breaking terms:
\begin{eqnarray}
V_{SOFT}=&=&{m_{1}^{0}}^{2}\mid H_{1}\mid ^{2}
+{m_{2}^{0}}^{2}\mid H_{2}\mid ^{2} + {m_{S}^{0}}^{2}\mid S\mid
^{2}+{m_{U}^{0}}^{2}\mid  {\tilde{t}}_{L}^{c} \mid^{2}\nonumber\\ &+& 
{m_{Q}^{0}}^{2}\mid {\tilde{Q}} \mid ^{2}- 
h_{s}^{0}A_{s}^{0}(S H_{1}.H_{2} + 
h. c.) - h_{t}^{0}A_{t}^{0}({{\tilde{t}}_{L}^{c*}}{\tilde{Q}}.H_{2} + 
h. c.)\nonumber\\&+& \sum_{a} M^{0}_{a} \lambda^{a}\lambda^{a}.
\end{eqnarray} 
Here $\lambda^{a}$ are gauginos with the masses $M^{0}_{a}$, and  
${\tilde{t}}_{L}^{c}$ and $\tilde{Q}$ are the scalar components of 
$\hat{U}^{c}$ and $\hat{Q}$, respectively. The superscript nought on each 
quantity designates its value at the SUGRA scale which is determined by the 
VEV's of the hidden sector fields (moduli and dilaton fields, in the 
case of string compactifications), modulo the threshold corrections.
  
The scalar components of the Higgs superfields are assigned the following 
representation under $SU(2)$ group: 
\begin{eqnarray}
\hat{H_{1}}\longrightarrow H_{1}=\left(\begin{array}{c c} 
H_{1}^{0}\\H_{1}^{-}\end{array}\right)\; , \; 
\hat{H_{2}}\longrightarrow H_{2}=\left(\begin{array}{c c} 
H_{2}^{+}\\H_{2}^{0}\end{array}\right)\; , \; 
\hat{S}\longrightarrow S\, .
\end{eqnarray}
Adding to $V_{SOFT}$ the usual F-term and D-terms of the associated  
group factors, one obtains the full scalar potential, whose  Higgs part 
reads 
\begin{eqnarray}
V_{0}&=&{m_{1}^{0}}^{2}\mid H_{1}\mid ^{2} 
+{m_{2}^{0}}^{2}\mid H_{2}\mid ^{2} + {m_{S}^{0}}^{2}\mid S\mid 
^{2}+ \lambda^{0}_{1}\mid H_{1}\mid ^{4} +\lambda^{0}_{2}\mid 
H_{2}\mid ^{4}\nonumber\\&+&\lambda^{0}_{S}\mid S\mid ^{4}+
\lambda^{0}_{12} \mid H_{1}\mid ^{2} \mid H_{2}\mid ^{2} +  
\lambda^{0}_{1S} \mid H_{1}\mid ^{2} \mid S\mid ^{2} + 
\lambda^{0}_{2S} \mid H_{2}\mid ^{2} \mid S\mid^{2}\nonumber\\&-& 
h_{s}^{0}A_{s}^{0}(S H_{1}.H_{2} + h. c.)\} 
\end{eqnarray}  
The terms involving parameters $\lambda_{i}^{0}$ come from the 
supersymmetric part of the Lagrangian consisting of $F$ and $D$ terms, 
and their explicit expressions are listed below: 
\begin{eqnarray}
\lambda_{1}^{0}&=&\frac{1}{8}G^{2}_{0}
+\frac{1}{2}{g_{Y'}^{0}}^{2}{Q'}_{1}^{2}\nonumber\\
\lambda_{2}^{0}&=&\frac{1}{8}G^{2}_{0}
+\frac{1}{2}{g_{Y'}^{0}}^{2}{Q'}_{2}^{2}\nonumber\\
\lambda_{S}^{0}&=&\frac{1}{2}{g_{Y'}^{0}}^{2}{Q'}_{2}^{2}\\ 
\lambda_{12}^{0}&=&-\frac{1}{4}G^{2}_{0}
+{g_{Y'}^{0}}^{2}{Q'}_{1}{Q'}_{2}+{h_{s}^{0}}^{2}\nonumber\\
\lambda_{1S}^{0}&=&{g_{Y'}^{0}}^{2}{Q'}_{1}{Q'}_{S}+ 
{h_{s}^{0}}^{2}\nonumber\\
\lambda_{2S}^{0}&=&{g_{Y'}^{0}}^{2}{Q'}_{2}{Q'}_{S}+
{h_{s}^{0}}^{2}\nonumber
\end{eqnarray}
where $G^{0}=\sqrt{{g^{0}_{2}}^{2}+{g^{0}_{Y}}^{2}}$. 
As there is no experimental constraint on $U(1)_{Y'}$ group, $g_{Y'}$ is 
arbitrary. However, for realistic models it is expected that $g_{1'}\sim 
g_{1}$ \cite{haber-sher}, so we assume the unification of $g_{1'}$ 
with the other gauge couplings at the the MSSM unification scale $M_{U}$ 
\cite{lang-uni}. Using the trace formulas for the fermion sector  
\begin{eqnarray} 
Tr[Q_{color}^{2}]=Tr[Q_{isospin}^{2}]=2\,,\, 
Tr[Y^{2}]=\frac{10}{3}\,,\,\nonumber\\ 
Tr[Y'^{2}]= 6{Q'}_{Q}^{2}+3({Q'}_{U}^{2}+{Q'}_{D}^{2})+ 
2{Q'}_{L}^{2}+{Q'}_{E}^{2} 
\end{eqnarray}
we normalize the gauge couplings such that, at $M_{U}$, they satisfy 
\begin{eqnarray}
g^{0}_{3}=g^{0}_{2}=g^{0}_{1}=g^{0}_{1'}=g^{0}\; ,
\end{eqnarray}
with the normalized $U(1)_{Y}$ and $U(1)_{Y'}$ couplings,
\begin{eqnarray}
g^{0}_{1}=\sqrt{\frac{5}{3}}g^{0}_{Y}\; ,\; 
g^{0}_{1'}=\sqrt{\frac{6{Q'}_{Q}^{2}+3({Q'}_{U}^{2}+{Q'}_{D}^{2}) 
+2{Q'}_{L}^{2}+ {Q'}_{E}^{2}}{2}}g^{0}_{Y'}\; . 
\end{eqnarray}

In obtaining the renormalization group flow of the parameters of the 
potential we shall consider one-loop RGE's which were listed in Appendix A 
of \cite{Biz}. We assume that the scale of SUSY breaking is around 
the weak scale, and thus we integrate RGE's of a softly broken SUSY model 
from the SUGRA scale down to the weak scale directly. Among the RGE's the 
most complicated ones are those involving Yukawa couplings $h_{s}$ and 
$h_{t}$ which obey coupled nonlinear equations. The top Yukawa coupling 
$h_{t}$ reaches its fixed point value of $h_{t}\sim 1 - 1.2$ almost  
independently of the initial conditions $h_{s}^{0}$ and $h_{t}^{0}$. 
Corresponding to this $h_{s}$ takes values around $0.6-0.8$. On the other 
hand, the RGE's of soft masses, being linear, can be solved exactly as a 
function of their initial conditions, for given $h_{s}^{0}$ and
$h_{t}^{0}$. Finally, as a by-product of the coupling constant unification, 
it is natural to assume a common mass $M_{1/2}$ for all gauginos at the 
SUGRA scale.

In constructing the solutions of RGE's one needs to specify the 
$U(1)_{Y'}$ charges of the fields. Without referring to specific $E(6)$ 
based charge assignments, one can relate different $U(1)_{Y'}$ charges to 
each other by imposing the cancellation of the triangular anomalies 
together with the gauge invariance of the potential. The superpotential 
in (1) includes only the top and Higgs trilinear mass terms, and thus we 
shall require the gauge invariance for these vertices only, leaving 
other the would-be vertices (such as $\hat{E}^{c}\hat{L}.\hat{H}_{1}$) 
unconstrained. Then the solution of $U(1)_{Y'}$ charges reads
\begin{eqnarray}
&&Q'_{S}=-(Q'_{1}+Q'_{2})\; , \; Q'_{U}=(Q'_{1}-3Q'_{2})/3\; , \;
Q'_{D}=(Q'_{1}+3Q'_{2})/3\,,\nonumber\\ &&Q'_{E}=-(Q'_{1}-Q'_{2})\;, \;
Q'_{Q}=-Q'_{1}/3\;,\; Q'_{L}=-Q'_{2},
\end{eqnarray}
which fix all but the two ($Q'_{1}$ and $Q'_{2}$) of the $U(1)_{Y'}$ 
charges. The advantageous side of this solution set is that it leaves the 
$U(1)_{Y'}$ charges of Higgs doublets free, which will be important in 
analyzing the mixing angle of $Z$ boson and $U(1)_{Y'}$ gauge boson. We 
give this solution for the third family, and assume vanishing $U(1)_{Y'}$ 
charges for the first two families. This is allowed in non-geometrical 
string compactifications (such as free-fermionic models), where a given 
Higgs doublet couples only one family \cite{faraggi}. As will be 
discussed later on, third family coupling of $U(1)_{Y'}$ is important in 
analyzing the $Z'$ models with LEP constraints. Finally, for future use, 
we make the choice $Q'_{1} =Q'_{2}= -1$ in (9) which fix all of the 
$U(1)_{Y'}$ charges. Then, the solution of RGE's, with the initial 
values $h_{s}^{0}=h_{t}^{0}=\sqrt{2}g^{0}$ for Yukawa couplings, read as 
follows: 
\begin{eqnarray}
h_{s}&=&0.595\nonumber\\
h_{t}&=&1.028\nonumber\\
A_{s}&=&0.42 A_{s}^{0} - 0.272 A_{t}^{0} - 0.285 M_{1/2}\nonumber\\
A_{t}&=&-0.045 A_{s}^{0} + 0.128 A_{t}^{0} + 1.755 M_{1/2}\nonumber\\
m_{1}^{2}&=&-0.064 {A_{s}^{0}}^{2} + 0.036 A_{s}^{0} A_{t}^{0} +
  0.007 {A_{t}^{0}}^{2}- 0.01 A_{s}^{0} M_{1/2} + 
  0.019 A_{t}^{0} M_{1/2}\nonumber\\ &+& 0.52 M_{1/2}^{2} + 
0.047 ({m_{Q}^{0}}^{2} 
  +{m_{U}^{0}}^{2}) - 0.16 {m_{S}^{0}}^{2}+ 0.84 
{m_{1}^{0}}^{2} 
  - 0.11 {m_{2}^{0}}^{2}\nonumber\\
m_{2}^{2}&=&-0.038 {A_{s}^{0}}^{2} + 0.037 A_{s}^{0} A_{t}^{0} -
  0.048 {A_{t}^{0}}^{2}+ 0.045 A_{s}^{0} M_{1/2} -
  0.19 A_{t}^{0} M_{1/2}\nonumber\\ &-& 2.47 M_{1/2}^{2}- 0.41
({m_{Q}^{0}}^{2} 
  +{m_{U}^{0}}^{2}) - 0.1 {m_{S}^{0}}^{2}- 0.1 {m_{1}^{0}}^{2}
  + 0.485 {m_{2}^{0}}^{2}\\
m_{S}^{2}&=&-0.128 {A_{s}^{0}}^{2} + 0.072 A_{s}^{0} A_{t}^{0} +
  0.014 {A_{t}^{0}}^{2} - 0.021 A_{s}^{0} M_{1/2} +
  0.039 A_{t}^{0} M_{1/2}\nonumber\\ &+& 0.081 M_{1/2}^{2}+
0.094 ({m_{Q}^{0}}^{2}
  +{m_{U}^{0}}^{2}) + 0.68 {m_{S}^{0}}^{2}- 0.32 
{m_{1}^{0}}^{2}
  - 0.22 {m_{2}^{0}}^{2}\nonumber\\
m_{U}^{2}&=&0.017 {A_{s}^{0}}^{2} + 0.0005 A_{s}^{0} A_{t}^{0} -
  0.037 {A_{t}^{0}}^{2} + 0.037 A_{s}^{0} M_{1/2} -
  0.139 A_{t}^{0} M_{1/2}\nonumber\\ &+& 3.27  M_{1/2}^{2}-
  0.306 {m_{Q}^{0}}^{2} + 0.69 {m_{U}^{0}}^{2} + 0.038 {m_{S}^{0}}^{2} 
  + 0.038 {m_{1}^{0}}^{2}- 0.27 {m_{2}^{0}}^{2}\nonumber\\
m_{Q}^{2}&=&0.009 {A_{s}^{0}}^{2} + 0.0002 A_{s}^{0} A_{t}^{0} -
  0.018 {A_{t}^{0}}^{2} + 0.019 A_{s}^{0} M_{1/2} -
  0.07 A_{t}^{0} M_{1/2}\nonumber\\ &+& 4.72  M_{1/2}^{2}
  +0.85 {m_{Q}^{0}}^{2} - 0.15 {m_{U}^{0}}^{2} + 0.02 {m_{S}^{0}}^{2}
  + 0.02 {m_{1}^{0}}^{2}- 0.13 {m_{2}^{0}}^{2}\nonumber
\end{eqnarray}
As a result of the normalization of the gauge couplings the low energy 
parameters are not very sensitive to the assignmets of the $U(1)_{Y'}$ 
charges. For example, if one chooses a model with $E(6)$ charge 
assignments, the results are affected only by a few percents. 
Thus, from a practical point of view, one can regard the above-listed 
solutions as independent of $U(1)_{Y'}$ charge assignments. On the 
contrary, dependence of the low energy parameters, especially the triliear 
couplings, on the variations of the initial conditions of Yukawa 
couplings  is important. In what follows we shall confine ourselves to 
$h_{s}^{0}=h_{t}^{0}=\sqrt{2}g^{0}$ \cite{faraggi}, as was already used 
in obtaining (10). The low energy potential with the parameters in (10) 
is a general one, and thus one has to specify the appropriate region of 
the parameter space to satisfy the phenomenological requirements 
existing at the weak scale. 
  
After the breaking of gauge symmetry down to $SU(3)_{c}\times 
U(1)_{em}$, there will arise two neutral massive gauge bosons $Z$ and 
$Z'$ whose mass matrix reads
\begin{eqnarray}
({\cal{M}}^{2})_{Z-Z'}=\left (\begin{array}{c c}
M_{Z}^{2}&\Delta^{2}\\\Delta^{2}&M_{Z'}^{2}\end{array}\right),
\end{eqnarray}  
where
\begin{eqnarray}
M_{Z}^{2}&=&\frac{1}{4}G^2(v_{1}^2+v_{2}^2),\\
M_{Z'}^{2}&=&{g}_{Y'}^{2}(v_{1}^{2}Q_{1}^{2}+v_{2}^{2}Q_{2}^{2}
+v_{s}^{2}Q_{S}^{2}),\\
\label{mix}
\Delta^{2}&=&\frac{1}{2}g_{Y'}\,G(v_{1}^2Q_{1}-v_{2}^2Q_{2}),
\end{eqnarray}
and, Higgs VEV's are defined as
\begin{eqnarray}
<H_{1}^{0}>=v_{1}/\sqrt{2}\; ; \; <H_{2}^{0}>=v_{2}/\sqrt{2}\; ; \; 
<S^{0}>=v_{s}/\sqrt{2}.
\end{eqnarray} 

There are three main conditions that the vacuum state must satisfy: 
\begin{itemize}
\item The $W$ boson mass must remain at its LEP1 value, as the model is
exteded only in the neutral direction,

\item The color and charge symmetries must remain unbroken,

\item The $Z-Z'$ mixing angle $\alpha$ must be below $a$ $few$ 
$\times 10^{-3}$ as otherwise the LEPI value of $M_{Z}$ is destructed.

\end{itemize}

The first condition can be satisfied using the fact that the potential (4)
has a common mass scale defined by the gravitino mass, as the soft SUSY 
breaking terms in (2) are generated by the SUGRA breaking. Thus, the mass 
scale of the potential, $m_{0}$ (proportional to the gravitino mass), can be 
factored out and remaining dimensionless potential can be minimized 
freely. The first condition above can then be met by imposing the constraint
\begin{eqnarray} 
m_{0}\sqrt{f_{1}^{2}+f_{2}^{2}}=246\,GeV
\end{eqnarray}
where $f_{1}$ and $f_{2}$ are the dimensionless $H_{1}$ and $H_{2}$ 
VEV's, which are defined as $v_{1}=m_{0}f_{1}, v_{2}=m_{0}f_{2}$, and for 
future use $v_{s}=m_{0}f_{s}$.

The charge breaking can arise from both Higgs and squark 
sectors. In the Higgs sector, with the help of  $SU(2)$ symmetry, a possible 
$<H_{2}^{+}>$ can be rotated away, and the charge breaking can be 
parametrised in terms of $v_{-}=<H_{1}^{-}>$. As can be calculated 
easily, the potential prefers the charge preserving minimum, if 
$\tilde{A}\, >\, 0$, where 
\begin{eqnarray}
\tilde{A}=2\sin^{2}\beta\, m_{H^{\pm}}^{2},
\end{eqnarray}
$\tan\beta=v_{2}/v_{1}$, and $m_{H^{\pm}}^{2}$ is the charged Higgs boson 
mass-squared. Consequently, in what follows we shall work in that 
portion of the parameter space where the charged Higgs boson has real 
mass, so that the charge breaking in the Higgs sector is avoided.

When, at least one of $<{\tilde{t}}_{L}^{c}>$, $<\tilde{Q}>$, take a 
nonzero value, both color and charge symmetries are broken. To prevent 
the formation of such a minimum, one has to have a certain 
hierarcy between the top trilinear coupling $A_{t}$ and the soft squark 
mass parameters. In fact, the usual criterium \cite{EGHRZ} 
$h_{t}^{2}A_{t}^{2} < 3(m_{Q}^{2} + m_{U}^{2} + m_{2}^{2})$ must hold at 
a scale $Q\sim A_{t}/h_{t}$. More importantly, when the squark masses go 
negative, a charge and/or color breaking minimum will be developed, even 
if it is secondary. When analyzing the low energy potential, we shall 
always keep these conditions in mind. 
 
That the $Z-Z'$ mixing angle is to be small is a severe constraint on the 
vacuum state. The $Z-Z'$ mixing angle, $\alpha$, is generated by 
the off-diagonal elements of the $Z-Z'$ mass matrix (10), and given by 
\begin{eqnarray}
\alpha=\frac{1}{2}\arctan\left(\frac{2\Delta^2}{M^{2}_{Z'}-M^{2}_{Z}}\right)
\end{eqnarray}

There are mainly three regions of parameter space of the pure Higgs sector 
yielding a small $\alpha$:
\begin{enumerate}
\item $Heavy$ $Z'$ $Minimum\; :$ 
As is seen from (18),  when $M_{Z'}>>M_{Z}$, $\alpha$ becomes small. This 
occurs in that portion of the parameter space satisfying,    
\begin{eqnarray}
-m_{S}^{2} >> |m_{1}^{2}|\, ,\, |m_{2}^{2}|\, , \, (h_{s}A_{s})^{2}
\end{eqnarray}
for which Higgs VEV's behave like 
\begin{eqnarray}
v_{s}\sim \sqrt{-m_{S}^{2}/\lambda_{s}} >> v_{1}\, , \, v_{2}. 
\end{eqnarray}
This ordering of the VEV's make $\Delta^{2}$ small compared to $M_{Z'}^{2}$, 
whereby producing a small mixing angle. This mechanism does not require any 
relationship among the soft masses $m_{1}^{2}$ and  $m_{2}^{2}$, and
charges $Q'_{1}$ and $Q'_{2}$; it utilizes only the ordering in (19). 
$\tan\beta$, however, is closely related to the values of 
charges $Q'_{1}$, $Q'_{2}$ and the values of the soft masses
$m_{1}^{2}$ and $m_{2}^{2}$. When $-m_{S}^{2}$ takes higher and higher 
values, only $U(1)_{Y'}$ gets broken, and other group factors 
remain unbroken. This introduces a mass scale $Q=\sqrt{-m_{S}^{2}}$ 
between weak and SUGRA scales; remnants of the gauge group must be 
broken by a second stage around the weak scale. Before the occurance of 
such an intermediate scale, large $-m_{S}^{2}$ can create a relatively 
large $\mu$ parameter, thus $-m_{S}^{2}$ getting large values must be 
avoided.

\item $Light$ $Z'$ $Minimum\; :$
As (18) suggests, another way of obtaining a small mixing angle would be 
to make $\Delta^{2}$ small irrespective of the value of $M_{Z'}$. To 
obtain this kind of cancellation, one needs roughly $v_{1}^{2}/v_{2}^{2} 
\sim |Q'_{2}/Q'_{1}|$ which may be satisfied in some limited portion of the 
parameter space. However, unlike MSSM and like NMSSM, the existence of the 
trilinear coupling parameter $h_{S}A_{s}$ opens a new avenue in obtaining an 
appropriate electroweak breaking. That is, when 
\begin{eqnarray}
h^{2}_{S}A_{s}^{2} >> |m_{1}^{2}|, |m_{2}^{2}|, |m_{S}^{2}|
\end{eqnarray}
holds, all VEV's are drawn approximately to the same point:
\begin{eqnarray}
v_{1}\sim v_{2}\sim v_{s} \sim A_{s}/(h_{S}\sqrt{2}). 
\end{eqnarray}
If the $U(1)_{Y'}$ charges of Higgs doublets satisfy $Q'_{2}=Q'_{1}$, 
this large trilinear coupling-induced minimum cancels $Z-Z'$ mixing, 
allowing $M_{Z}$ to be compatible with LEPI data. Such a light $Z'$ must 
show up in the $Z$-pole observables, and therefore parameters of the model 
can be constrained using LEP1 data. As a side remark, the passage of the 
trilinear coupling from lower values to higher ones (compared to the soft 
masses) is either a first order or second order phase transition,  
depending on the sign of $m^2=m_{1}^{2}+m_{2}^{2}+m_{S}^{2}$. In fact, 
if $m^{2}$ is positive, there is a first order phase transition 
occuring at the critical point $A_{s}^{crit}=\sqrt{8\,m^{2}/3}$.  

\item $Hybrid$ $Minimum\; :$
One can obtain a phenomenologically viable minimum in all aspects by 
combining the parameter spaces of $Heavy$ $Z'$ $Minimum$ and $Light$ $Z'$ 
$Minimum$. To do this, one can choose both $-m_{S}^{2}$ and $A_{s}^{2}$ 
large compared to other soft masses. As large $A_{s}^{2}$ pushes all 
Higgs VEV's to the same point, and $-m_{S}^{2}$ prefers a large 
SM-singlet VEV, when both $-m_{S}^{2}$ and $A_{s}^{2}$ are large,
doublet VEV's (controlled by $A_{s}$ via (22)) will approach to 
approximately the same value, and SM-singlet VEV (controlled by 
$-m_{S}^{2}$ via (20)) will be much larger than them. Thus, this 
portion of the parameter space yields a small mixing angle together with a 
relatively heavy $Z'$. In the analyses below, we shall be mainly 
interested in this kind of parameter space and assume $Q'_{1}=Q'_{2}$. 
Actually, when $-m_{S}^{2}$ is large, one does not need this condition, 
as $\Delta^{2}$ in (14) is already small compared to $M_{Z'}^{2}$. 
However, this kind of choice allows for the realization of the required 
minimum in a reasonable range of parameter values. 
\end{enumerate}

In the above- mentioned low- energy analysis we have required a certain 
hierarchy among the potential parameters. However, as dictated by the 
solution of RGE's in (10), it is not realistic to consider such idealized 
cases since as one parameter changes, all others do too as a function of 
the initial conditions. Moreover, we need not only the Higgs sector 
parameters, bu also the parameters of the squark sector as we shall  
calculate the 1-loop squark contributions to the Higgs potential. Thus, 
one has to determine the SUGRA scale parameter space consistently by 
considering all the low energy parameters simultaneously.  
 
Once non-universality is permitted, one faces with a huge parameter space 
each point of which corresponds to some symmetry breaking scheme at low 
energies. The usual procedure for the determination of the appropriate 
portion of the parameter space, would be to trace the SUGRA-scale parmeter 
space point-by-point, and pick up those yielding a viable minimum at  
low energies. However, with the low energy parameters in (10), and the 
constraints implied by the $Hybrid$ $Minimum$, we can determine the 
appropriate parameter space analytically. In accordance with the 
conditions coming from the $Hybrid$ $Mininum$, we want to speed up 
$-m_{S}^{2}$ and $A_{s}$ compared to other mass parameters in terms of 
their dependence on a choosen SUGRA scale parameter. These two have three 
parameters, $A_{s}^{0}$, $A_{t}^{0}$ and $M_{1/2}$, in common, and we 
shall use $A_{s}^{0}$ as a probe for determining the appropriate 
parameter space, by fixing other parameters with minimal amount of 
non-universality \cite{MN}. To slow down the $A_{s}^{0}$ dependence of 
$m_{1}^{2}$ and $m_{2}^{2}$, we make the following choice for 
$m_{1}^{0\, 2}$ and $m_{2}^{0\, 2}$: 
\begin{eqnarray}
m_{1}^{0\,2}\approx m_{2}^{0\,2}\approx A_{s}^{0\,2}/10
\end{eqnarray}
For the remaining parameters we keep universality:
\begin{eqnarray}
m_{S}^{0\,2}=m_{Q}^{0\,2}=m_{U}^{0\,2}=A_{t}^{0\,2}/3,
\end{eqnarray} 
and finally we let $M_{1/2}=A_{t}^{0}$. Under these conditions, the low 
energy parameters read as follows: 
\begin{eqnarray}
A_{s}&\approx & 0.42\, A_{s}^{0}-0.58\, A_{t}^{0}\nonumber\\
A_{t}&\approx & -0.045\, A_{s}^{0}+1.9\, A_{t}^{0}\nonumber\\
m_{1}^{2}&\approx &0.026\, A_{s}^{0}A_{t}^{0} + 0.53\, 
A_{t}^{0\, 2}\nonumber\\
m_{2}^{2}&\approx &0.08\, A_{s}^{0}A_{t}^{0} -3.5\, 
A_{t}^{0\, 2}\nonumber\\
m_{S}^{2}&\approx &-0.18\, A_{s}^{0\, 2}+0.05\, A_{s}^{0}A_{t}^{0} +0.43\,  
A_{t}^{0\, 2}\\
m_{Q}^{2}&\approx &-0.0026\, A_{s}^{0\, 2}+0.02\, A_{s}^{0}A_{t}^{0} +5.0\,
A_{t}^{0\, 2}\nonumber\\
m_{U}^{2}&\approx &-0.0052\, A_{s}^{0\, 2}+0.04\, A_{s}^{0}A_{t}^{0} 
+3.4\, A_{t}^{0\, 2}\nonumber
\end{eqnarray}
As we see from these equations the quadratic $A_{s}^{0\, 2}$ dependence of 
$m_{1}^{2}$ and $m_{2}^{2}$ are supressed compared to those of the others 
with $m_{S}^{2}$ being the fastest one among all concerning their 
$A_{s}^{0\, 2}$ dependence. In the analysis below we shall vary 
\begin{eqnarray}
\zeta=A_{s}^{0}/A_{t}^{0}
\end{eqnarray}
in a certain range of values, and determine $A_{t}^{0}$ from the invariance 
of the $W^{\pm}$ mass under such Abelian extensions of MSSM. When  
$\zeta$ is small, except for $m_{2}^{2}$, all soft masses are positive, 
yielding a non-zero $f_{2}$ and vanishing $f_{1}$, $f_{2}$. This is not an 
acceptable minimum, as the gauge symmetry is not broken completely. As 
$\zeta$ increases, $m_{2}^{2}$ starts overcoming the large negative 
threshold coming from (essentially the $SU(3)_{c}$) gaugino masses, and 
$H_{1}^{0}$ and $S$ do develop nonzero VEV's. But still it may not yield 
a small enough mixing angle. Further increase of $\zeta$ brings us to 
the sought minimum where $0 \neq f_{1} \sim f_{2} << f_{s}$. However, 
this increase cannot be maintained further as the squark masses turn to 
negative after overcoming their large positive mass thresholds dominated  
by the $SU(3)_{c}$ gaugino. Negative squark masses cause charge and color 
breaking minima, even if secondary, so that this limiting case will be 
avoided below. 
\section{One-Loop Corrections}
Until now our discussion has been based solely on the RGE-improved 
tree-level potential. However, quantum corrections beyond the log effects 
included in the RGE analysis are important. Especially the top-stop 
sector gives the most important contribution due to the relatively large 
value of top Yukawa coupling. To take such radiative corrections into 
account we shall follow efective potential approach \cite{CQW,QUIROS} 
in which the radiatively corrected one-loop potential is given by 
\begin{eqnarray}
V_{1}=V+\Delta V
\end{eqnarray}
where the one-loop contribution has the Coleman-Weinberg form
\begin{eqnarray}
\Delta V = \frac{1}{64\pi^2} 
Str{\cal{M}}^{4}\ln\frac{{\cal{M}}^{2}}{Q^2}\; ,
\end{eqnarray}
where $Str$ is the usual supertrace and ${\cal{M}}^{2}$ is the field 
dependent mass-squared matrix. We have transferred a renormalization 
scheme dependent constant into a redefinition of the renormalization 
scale $Q^{2}$. One notes that all the parameters in (27) are to be 
evaluated at the scale $Q^{2} \sim {\cal{O}}(v_{1}^{2}+v_{2}^{2})$. 
Indeed, this is consistent with the RGE analysis of the last section as we 
have integrated them from the SUGRA scale down to the weak scale. 
In the loop expression (28) we consider only the contributions of top 
and stops $\tilde{t}_{2}$, $\tilde{t}_{1}$ whose masses are given by
\begin{eqnarray}
m_{t}=h_{t}|H_{2}^{0}| \; ; \;
m^{2}_{\tilde{t}_{1,2}}=\frac{1}{2}\{m_{11}^{2}+m_{22}^{2}\pm  
\sqrt{(m_{11}-m_{22})^{2}+4 |m_{12}|^{2}}\} 
\end{eqnarray}
where 
\begin{eqnarray}
m_{11}&=&m_{Q}^{2}+\lambda_{1t}|H_{1}|^{2}+\lambda_{2t}|H_{2}|^{2}+ 
\lambda_{st}|S|^{2}+\tilde{\lambda}_{1Q}|H_{1}^{0}|^{2}+ 
\tilde{\lambda}_{2Q}|H_{2}^{+}|^{2}\nonumber\\
m_{22}&=&m_{U}^{2}+\lambda_{1u}|H_{1}|^{2}+\lambda_{2u}|H_{2}|^{2}+ 
\lambda_{su}|S|^{2}\\
m_{12}&=&-h_{t}A_{t}H_{2}^{0\, *}+h_{s}h_{t}S^{*}H_{1}^{0\, *}\nonumber
\end{eqnarray} 
and the dimensionless coefficients 
\begin{eqnarray}
\lambda_{1t}&=&-g_{2}^{2}/4-g_{Y}^{2}/12+ 
g_{Y'}^{2}Q'_{1}Q'_{Q}\nonumber\\
\lambda_{2t}&=&-g_{2}^{2}/4-g_{Y}^{2}/12+
g_{Y'}^{2}Q'_{2}Q'_{Q}+h_{t}^{2}\nonumber\\
\lambda_{st}&=&g_{Y'}^{2}Q'_{S}Q'_{Q}\nonumber\\
\lambda_{1u}&=&g_{Y'}^{2}Q'_{1}Q'_{U}\\
\lambda_{2u}&=&g_{Y'}^{2}Q'_{2}Q'_{U}+h_{t}^{2}\nonumber\\
\lambda_{su}&=&g_{Y'}^{2}Q'_{S}Q'_{U}\nonumber\\
\tilde{\lambda}_{1Q}&=&g_{2}^{2}/2\nonumber\\
\tilde{\lambda}_{2Q}&=&g_{2}^{2}/2-h_{t}^{2}\nonumber
\end{eqnarray}
follow from the colored sector of the full scalar potential. We shall 
calculate radiative corrections to the lightest CP-even Higgs mass 
bound, so we are interested in the CP-even scalar mass-squared matrix which 
can be obtained by evaluating $\frac{\partial^{2} V_{1}}{\partial \phi_{i} 
\partial \phi_{j}}$ at the VEV's, in the basis ($Re[H_{1}^{0}],
Re[H_{1}^{0}], Re[S^{0}]$).

Before going into a detailed numerical analysis, we first present a
general discussion on the effects of the radiative corrections based on 
some approximate formulae. The top-stop splitting $S_{t\tilde{t}}= 
\ln\frac{m_{\tilde{t}_{1}}m_{\tilde{t}_{2}}}{m_{t}^{2}}$ and stop 
splitting $S_{\tilde{t}\tilde{t}}=m^{2}_{\tilde{t}_{1}}-m^{2}_{\tilde{t}_ 
{2}}$ describe the most important contributions of the one-loop  
corrections. On the other hand, due to the choice of $Q^{2}$, the 
remaining log $\ln\frac{m_{t}^{2}}{Q^{2}}$ is not as important as the 
former ones. To extract some information about the effects of the loop 
corrections on the tree level parameters, one can expand the 
minimization equations in powers of stop splitting and identify the 
renormalization effects on the tree-level quantities. In fact, to lowest 
order in stop splitting $S_{\tilde{t}\tilde{t}}$, and neglecting the 
terms involving the gauge couplings, one finds that the most important 
contributions come to $A_{s}$, $m_{2}^{2}$ and $\lambda_{2}$; which are 
given by 
\begin{eqnarray}
\hat{A}_{s}&=&A_{s}+\beta_{h_{t}}S_{t\tilde{t}}A_{t}\nonumber\\
\hat{m}_{2}^{2}&=&m_{2}^{2}+\beta_{h_{t}}[ 
(A^{2}+A_{t}^{2})S_{t\tilde{t}}-A^{2}]\\
\hat{\lambda}_{2}&=&\lambda_{2}+\beta_{h_{t}}S_{t\tilde{t}}h_{t}^{2}\nonumber
\end{eqnarray}
where $\beta_{h_{t}}=\frac{3}{(4\pi)^{2}}h_{t}^{2}$, and 
$A^{2}=m_{Q}^{2}+m_{U}^{2}$. Let us now discuss the implications of these 
one-loop corrections in the light of the RGE solution set in (25). As the 
first equation in (32) shows, $A_{s}$ is strengthened by the loop 
corrections. However, $A_{s}^{0}$ dependence of $A_{t}$ is approximately 
one order of magnitude smaller than that of $A_{s}$ so that one  does  
not have a significant improvment for $A_{s}$, unless top-stop splitting 
is large. The improvement in $m_{2}^{2}$, $\delta m_{2}^{2} \sim 
\beta_{h_{t}}A^{2}(S_{t\tilde{t}}-1)$ depends, in addition to $A^{2}$ 
itself, on how large $S_{t\tilde{t}}$ is compared to unity. That is, if the 
stop masses are large compared to $m_{t}$, the top-stop splitting can be 
large enough to give a significant contribution to $m_{2}^{2}$. Hence, 
both $A_{s}$ and $m_{2}^{2}$ get significantly improved if the top-stop 
splitting is large enough. As one can read off from (25), for small 
$\zeta$, the squark soft masses are large due to the contribution 
of the $SU(3)_{c}$ gaugino. Therefore, $m_{2}^{2}$ is significantly 
improved by the loop corrections in this range of the $\zeta$ 
values. However, we observe from (25) that for larger $\zeta$ values
$A^{2}$ gets smaller and the loop contributions drop significantly. 
Finally, as dictated by (32), $H_{2}$ quartic coupling $\lambda_{2}$ is 
significantly improved by the radiative corrections if the top-stop 
splitting is large enough. In the small $\zeta$ limit, one has 
$f_{2}\sim \sqrt{-m_{2}^{2}/\lambda_{2}} >> f_{1}, f_{s}$, which 
clearly shows that tree-level $f_{2}$ is larger than the loop-level one. 
This radiative reduction in $f_{2}$ causes one-loop $\tan\beta$ to drop:
\begin{eqnarray}
\tan^{2}\hat{\beta}= \tan^{2}\beta (1-\tan\beta\, 
\frac{\beta_{h_{t}}[(A^{2}+A_{t}^{2}+2m_{t}^{2})S_{t\tilde{t}}-A^{2}]
}{A_{s}\mu_{s}})
\end{eqnarray}
where the effective MSSM $\mu$ parameter $\mu_{s}=(h_{s}v_{s})/\sqrt{2}$ 
is introduced. Indeed, with the contributions of $\hat{m}_{2}^{2}$ and 
$\hat{\lambda}_{2}$, one-loop $\tan\beta$ is reduced compared to the 
tree-level one. However, the inverse $\mu_{s}A_{s}$ dependence will 
force the loop contribution to drop rapidly after some $\zeta$ values. 
In this sense, one expects tree- and loop-level $\tan\beta$'s be close to 
each other in the large $\zeta$ regime. 
 
At the tree-level, for small $\zeta$, one expects a relatively large 
mixing angle as the expected cancellation in $\Delta^{2}$ does not 
occur. At the loop-level, however, due to the reduced $\tan\beta$
one expects a smaller $Z-Z'$ mixing angle as can be seen from the form of 
$\Delta^{2}$ in (14). Altough in the large $\zeta$ limit radiative 
correction to $\tan\beta$ is diminished, due to the increase in $v_{s}$ 
($-m_{S}^{2}$ increases) the loop-level $Z-Z'$ mixing angle will be still 
smaller than the tree-level one.
     
The lightest CP-even Higgs mass has the tree-level bound of
\begin{eqnarray}
m_{h_{1}}^{2\, max}= M_{Z}^{2}\cos^{2}2\beta + 
(v_{1}^{2}+v_{2}^{2})[\frac{h_{s}^{2}\sin^{2}2\beta}{2} + 
g_{Y'}^{2}(Q'_{1}\cos^{2}\beta +Q'_{2}\sin^{2}\beta)]\; . 
\end{eqnarray}
Here the first term is the MSSM tree level bound, the $h_{s}^{2}$ term is 
the NMSSM contribution ($g_{Y'}=0$ case), and finally the last term is the 
D-term contribution of $U(1)_{Y'}$ group. Using the same approximations 
that had lead us to (32), a straightforward calculation yields the 
following one-loop bound:
\begin{eqnarray}
\hat{m}_{h_{1}}^{2\, max}= m_{h_{1}}^{2\, 
max}+\beta_{h_{t}}[(\mu_{s}\cos\beta+A_{t}\sin\beta)^{2}+4 
S_{t\tilde{t}}m_{t}^{2}]\; .
\end{eqnarray} 
As we observe from this equation, one-loop bound is always larger than 
the tree-level one. Since $A_{s}^{0}$ dependence of $A_{t}$ is weak, the 
main contribution to the bound comes from $m_{t}$ and $\mu_{s}$ terms, and 
it is maximized either by top-stop splitting contribution, or by 
the $\mu_{s}$ contribution. In fact, due to this $\mu_{s}$ dependence it 
will be much larger in $Heavy$ $Z'$ $Minimum$ than in $Light$ $Z'$ $Minimum$.
In the parameter space we shall trace one expects the one-loop bound be 
dominated by $m_{t}$ and $\mu_{s}$ terms in the small and large $\zeta$ 
regimes, respectively.

Had we included the entire particle spectrum and worked to all orders 
our results would be $Q^{2}$ independent. The one-loop expressions for 
$\hat{A}_{s}$ and $\hat{m}_{2}^{2}$ in (32) are actually $Q^{2}$ 
dependent and their dependence can be recovered by letting 
$S_{t\tilde{t}} \longrightarrow S_{t\tilde{t}} + \ln  
\frac{m_{t}^{2}}{Q^{2}}$. Therefore, if for some choice of $Q^{2}$, $\ln 
\frac{m_{t}^{2}}{Q^{2}}$ happens to be important, one can analyze 
these two quantities by intoducing the splitting function 
$S_{Q\tilde{t}}=\ln\frac{m_{\tilde{t}_{1}}m_{\tilde{t}_{2}}}{Q^{2}}$. 
Similarly, in the expression for $\tan\beta$, except for $m_{t}^{2}$ term, 
one can make the replacement $S_{t\tilde{t}} \longrightarrow 
S_{Q\tilde{t}}$ to take into account its $Q^{2}$ dependence. Unlike these, 
$\hat{\lambda}_{2}$ in (32), and lightest Higgs mass bound in (35) are 
independent of $Q^{2}$, so they exhibit the same behaviour for all $Q^{2}$. 
In general, all scalar mass-squared matrices are $Q^{2}$ dependent, but 
the lightest CP-even Higgs mass bound turns out to be scale independent. 
In the RGE analysis assumed that the scale of the SUSY breaking 
$M_{SUSY}$ is around the weak scale, so the choice of $Q^{2}\sim 
(v_{1}^{2}+v_{2}^{2})$ is necessary for consistency of the analysis. Thus, 
we do not expect the $Q^{2}$ dependence of the parameters to change the 
above-mentioned predictions significantly. 
 
Until now we have based our analysis on the approximate formulae 
which were obtained by assuming that the top-stop splitting, 
$\ln\frac{m_{t}^{2}}{Q^{2}}$, and all gauge couping dependent terms are 
negligably small. These assumptions do not necessarily hold in the entire 
$\zeta$ spectrum, and thus it is needed to have a detailed picture for the 
$\zeta$ dependence of all these quantities.

\section{Numerical Analysis}
In this section we shall investigate the effects of the radiative 
corrections on various quantities by an exact treatment of the problem 
using numerical techniques. To obtain the scalar mass matrices one 
has to calculate $\frac{\partial^{2} V_{1}}{\partial \phi_{i} \partial 
\phi_{j}}$ evaluated at the VEV's. During the minimization we shall rescale 
all fields and parameters of mass dimension by $A_{t}^{0}$; consequently 
the parameters of the potential depend on a single quantity, $\zeta$ 
defined in (26). After minimizing the dimensionless potential, we recover 
the physical shell by requiring 
\begin{eqnarray}
A_{t}^{0}=v/\sqrt{f_{1}^{2}+f_{2}^{2}}
\end{eqnarray}
where $v=246\; GeV$ is the Fermi scale. As we have already discussed 
in obtaining (16), this rescaling procedure works very well for the tree 
level potential \cite{Biz} due to the uniqueness of the mass scale.
However, radiative corrections do necessarily introduce an additional mass 
scale $Q^{2}$. Thus, the rescaling invariance of the tree-level potential 
does not hold at the loop level. Using (36), one would rescale the basic 
log in (28) as 
\begin{eqnarray}
\ln\frac{{\cal{M}}^{2}}{Q^2} = \ln 
{\tilde{\cal{M}}}^{2}\frac{A_{t}^{0\, 2}}{Q^2}
\end{eqnarray}
which clearly requires the knowledge of $A_{t}^{0}$ which itself is 
something we aim to find. The determination of $A_{t}^{0}$ thus requires a 
consistency analysis where one inserts a trial value in this rescaled 
log, and compare it with the resulting one after the minimization. This 
procedure goes on until trial and output values for $A_{t}^{0}$ do match. 
We have done this numerically, and the result is shown in Fig.1 as a 
function of $\zeta$.
 
In the analysis below we shall present tree-level and one-loop 
quantities on the same graph for the sake of easy comparison. In each 
graph the free variable is $\zeta$, the ratio of Higgs trilinear 
coupling to top trilinear coupling at the SUGRA level. The starting 
value of $\zeta$ is chosen to be that one for which none of the VEV's 
vanish. On the other hand, the maximum of $\zeta$ is determined by its 
threshold value at which $m^{2}_{\tilde{t}_{2}}$ turns to negative due 
to large $v_{s}$ values. This threshold shows up before squark soft masses 
turn to negative; so there is no danger of charge and/or color breaking in 
the range of $\zeta$ values we shall consider below.
 
Fig.1 shows the $\zeta$ dependence of $A_{t}^{0}$ for tree- and 
loop-level analyses. While tree-level $A_{t}^{0}$ peaks around $\zeta=7$ 
with a value $73\; GeV$, one-loop $A_{t}^{0}$ peaks around $\zeta=8$ 
hitting the value of $82\; GeV$. As the VEV's leave the small $\zeta$ 
regime (or $f_{2}$ dominated regime), the sum $f_{1}^{2}+f_{2}^{2}$ 
approach its minimum, as $f_{1}$ and  $f_{2}$ are driven to close 
enough values by $A_{s}$. That $A_{t}^{0}$ will be maximized around these 
$\zeta$ values follows from (36). After passing by this maximization 
point, all of the mass parameters become proportional to the associated 
power of $\zeta$, as dictated by (25). Hence, in this 'large $\zeta$' 
regime, we expect dimensionless doublet VEV's be approximately 
proportional to $\zeta$ because of which we observe an approximate 
$1/\zeta$ fall off in Fig.1. This kind of behaviour in $A_{t}^{0}$ 
reflects itself in all the relevant masses we shall discuss below.
In solving the RGE's we have used the prescription $M_{1/2}=A_{t}^{0}$, 
which implies the weak scale masses of $\sim 2.6\times A_{t}^{0}$, $0.8 
\times A_{t}^{0}$, $A_{t}^{0}/4$, and $A_{t}^{0}/10$ for  
$SU(3)_{c}$, $SU(2)$, $U(1)_{Y}$, and $U(1)_{Y'}$ gauginos, respectively. 
Thus, depending on the present and future experimental limits on the 
gaugino masses of different group factors, one can restrict $\zeta$ to 
a certain range of values, keeping in mind that the choice 
$M_{1/2}=A_{t}^{0}$ itself is not necessarily unique.
 
We plot the $\zeta$ dependence of $\tan\beta$ in Fig. 2. As we observe 
from this figure, for small $\zeta$ values, loop contributions do really 
push $\tan\beta$ to smaller values. Again in agreement with our 
expectations, for large  $\zeta$, both the tree level and loop results 
come closer rapidly, and gradually approach to unity . The difference 
between one-loop and tree-level $\tan\beta$'s fall below $1\%$ 
after $\zeta \sim 8$.

In Fig.3 we present the $\zeta$ dependence of the $Z-Z'$ mixing angle 
$\alpha$. In agreement with the predictions of the last section, 
one-loop mixing angle is smaller than the tree level one everywhere. As we 
see from this figure, the phenomenological bound of $\alpha_{max} \sim$ 
$a$ $few$ $\times$ $10^{-3}$ after $\zeta \sim 7$ is comfortably 
satisfied after $\zeta \sim 7$. The last point about this figure is 
that for large $\zeta$ values one-loop and tree-level results remain 
approximately parallel, indicating the fact that both doublet VEV's reach 
their limiting values controlled by $A_{s}$, and SM-singlet VEV enter the 
$-m_{S}^{2}$ dominated regime.
   
Another important quantity, $M_{Z_{2}}$, is shown in Fig. 4 as a function 
of $\zeta$. First, we see that loop corrections generally increase the 
$Z_{2}$ mass in the entire $\zeta$ range. Both tree-level and one-loop 
masses increase until $\zeta \sim 10$ in accordance with $A_{t}^{0}$ in 
Fig.1. Likewise, in parallel with the behaviour of $A_{t}^{0}$ 
$M_{Z_{2}}$ decreases gradually after $\zeta \sim 10$, and is expected 
to saturate after some point due to the fact that $\zeta$ dependence of 
$A_{t}^{0}$ and dimensionless SM-singlet VEV are almost inversely 
proportional to each other in this range of $\zeta$ values. The one-loop 
$M_{Z_{2}}$ peaks at $\zeta \sim 11$ by taking the value of $\sim 405\, 
GeV$; thus, it cannot increase indefinitely with $\zeta$. As expected, 
the tree-level $M_{Z_{2}}$, in similarity with the tree-level $A_{t}^{0}$,
peaks at $\zeta \sim 10$, with a value $\sim 330\; GeV$. The values taken 
by $M_{Z_{2}}$ depends crucially on the value of $g_{Y'}$. Under the 
normalization in (8), and the $U(1)_{Y'}$ charge assignments in (9), the 
solution of the RGE's yield $g_{Y'}/g_{Y}\approx 0.65$ at the weak scale. 
This ratio would push the value of $M_{Z_{2}}$ above $600\, GeV$, if 
$g_{Y'}=g_{Y}$ were the case. One notes that the recent Tevatron 
result \cite{tevatron} giving $M_{Z_{2}}\geq 590\, GeV$ would be well 
satisfied if $g_{Y'}$ were equal to $g_{Y}$.

As explained in the Introduction, one of the basic aims of constructing 
such extended models is of course the dynamical formation of the MSSM 
$\mu$ parameter. The effective $\mu$ parameter in the present model has 
the $\zeta$ dependence shown in Fig. 5, for which we have almost the 
same behaviour observed in $Z_{2}$ mass, as both are controlled by 
$A_{t}^{0}$. The one-loop $\mu_{s}$ peaks around $\zeta=11$, and takes 
the value $\sim 350\, GeV$ at this point. Similarly, the tree level 
$\mu_{s}$ peaks around $\zeta=10$ with a value $\sim 280\; GeV$.
  
Finally, in Fig. 6, we present the $\zeta$ dependence of the lightest 
Higgs mass bound, $m_{h_{1}}^{max}$, which is seen to satisfy the 
predictions of the last section. As we see from (34), the tree level 
bound depends solely on the doublet VEV's, so that after reaching the 
$\tan\beta \sim 1$ regime the bound is maximized and saturated at 
$m_{h_{1}}^{max} \sim 118\, GeV$. In the same way until leaving the small 
$\zeta$ region, one-loop bound also increases and hits the value $\sim 
125\, GeV$ in the far end of the total $\zeta$ range. The fact that 
one-loop bound saturates much later than the tree-level one is due to the 
$\mu_{s}$ dependence of the radiative corrections.

\section{Conclusions and Discussions}
In this work we have investigated one-loop contributions to certain 
low-energy quantities in the framework of the effective potential 
approach, using an RGE-improved radiatively corrected scalar potential, 
following from the superpotential in (1). However, derivation of the 
entire low-energy particle spectrum (such as top, bottom, and $\tau$ 
masses) requires the study of a more general superpotential involving, in 
addition to the superpotential in (1), exotics predicted in most string 
models and non-renormalizable quartic mass terms from which light 
fermion masses follow \cite{faraggi,jose}. Here we have restricted 
ourselves mainly to the study of certain low-energy quantities determined 
by the Higgs sector of the model, for which the typical superpotential 
in (1) should suffice \cite{Biz,jose}.
  
Among the low-energy quantities we have worked out, $Z'$ boson and lightest 
Higgs mass bound are of phenomenological importance. The 
search for $Z'$ \cite{rizzo} will be one of the goals of the next 
generation accelerators. In near future, LEP II will be searching for 
$Z'$ boson in leptonic and $WW$ channels. Besides this, LHC will search 
for $Z'$ boson with quark-antiquark fusion processes. In general, the 
exclusion limits of $Z'$ mass and its couplings depend on the model and 
collider parameters \cite{rrl}. The $U(1)_{Y'}$ charges of the present 
model are generation dependent, and thus, the constraints on its $Z'$ 
boson is weaker than that of the generation independent ones. The $Z-Z'$ 
mixing angle (See Fig. 3) is small enough to supress the effects of $Z'$ 
fermion couplings in the $Z$ -pole observables \cite{kolda}. The $Z'$ mass
in the present model has an upper bound of $\sim 400 \, GeV$, and 
satisfies the presently existing phenomenological bounds. 

The lightest Higgs mass bound turns out to be $\sim 125 \, GeV$ 
\cite{jose2} in MSSM. In NMSSM, however, it is $\sim 140 \, GeV$ 
\cite{EKW2-Ell}. In the present model, it turns out to be $\sim 125\, 
GeV$. The bounds of the present model and MSSM practically coincide, 
however, the bound in the present model is expected to increase 
slightly if the NNL corrections are taken into account \cite{jose2}.
In near future, the lightest Higgs boson will be discovered at LEP II if 
$m_{h_{1}}\leq 95\, GeV$, and at FNAL in $m_{h_{1}}\leq 120\, GeV$ after 
accumulating an integrated limunosity of 25-30 $fb^{-1}$ \cite{SMW}.

In conclusion, we have analyzed the efects of the radiative corrections 
on the various low energy quantities in the present model by taking the 
contributions of top and stops into account. As we have shown 
graphically, the one-loop improvment in the low energy parameters are 
no way negligable. Moreover, the one-loop corrections support 
the satisfaction of the phenomenological requirements compared to the 
bare tree-level potential. The findings of the work will be tested in the 
near-future colliders.
 
\section{Acknowledgements}
One of us (D. A. D.) thanks Paul Langacker for his helpful remarks at the 
earliest stages of this work.

\newpage
\begin{figure}
\vspace{10cm}  
\end{figure}  
\begin{figure} 
\vspace{12.0cm}
    \includegraphics{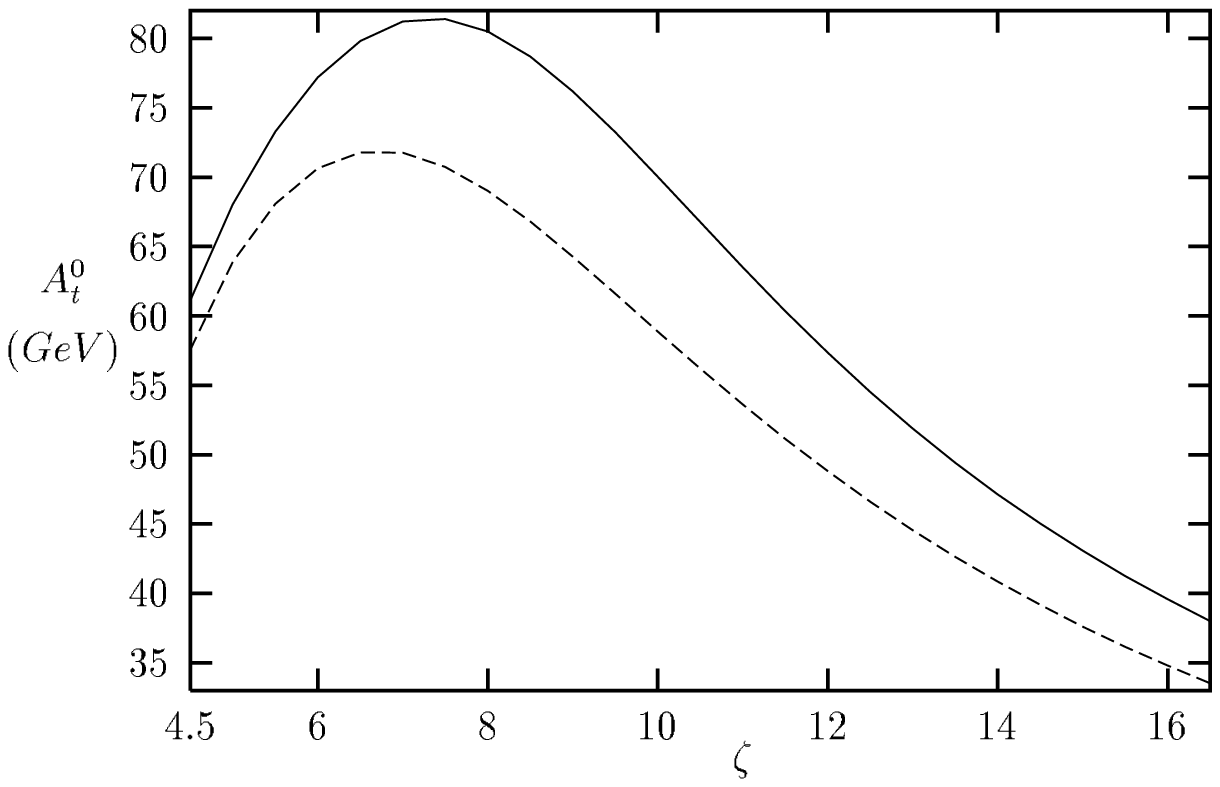}
    \vspace{-8.0cm}
\vspace{0.0cm}
\mbox{ \hspace{0.5cm} \large{\bf Figure 1: Dependence of $A_{t}^{0}$ on
$\zeta$ (solid line:}}
\mbox{ \hspace{0.5cm} \large{\bf one-loop, dashed line: tree-level)}}
\end{figure}
\begin{figure}
\vspace{12cm}
\end{figure}
\begin{figure}
\vspace{12.0cm}
    \includegraphics{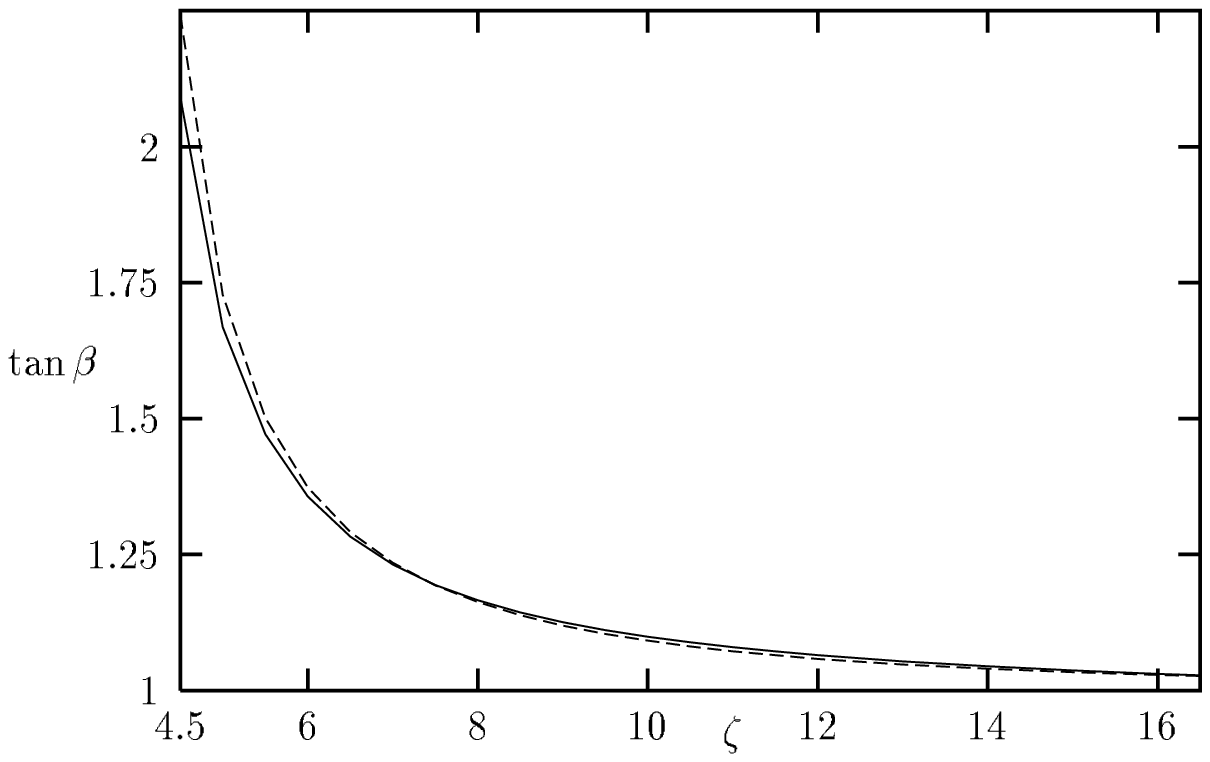}  
    \vspace{-8.0cm}
\vspace{0.0cm}
\mbox{ \hspace{0.5cm} \large{\bf Figure 2: Dependence of $\tan\beta$ on
$\zeta$ (solid line:}}
\mbox{ \hspace{0.5cm} \large{\bf one-loop, dashed line: tree-level)}} 
\end{figure}
\begin{figure}
\vspace{12cm}
\end{figure}
\begin{figure}
\vspace{12.0cm}
    \includegraphics{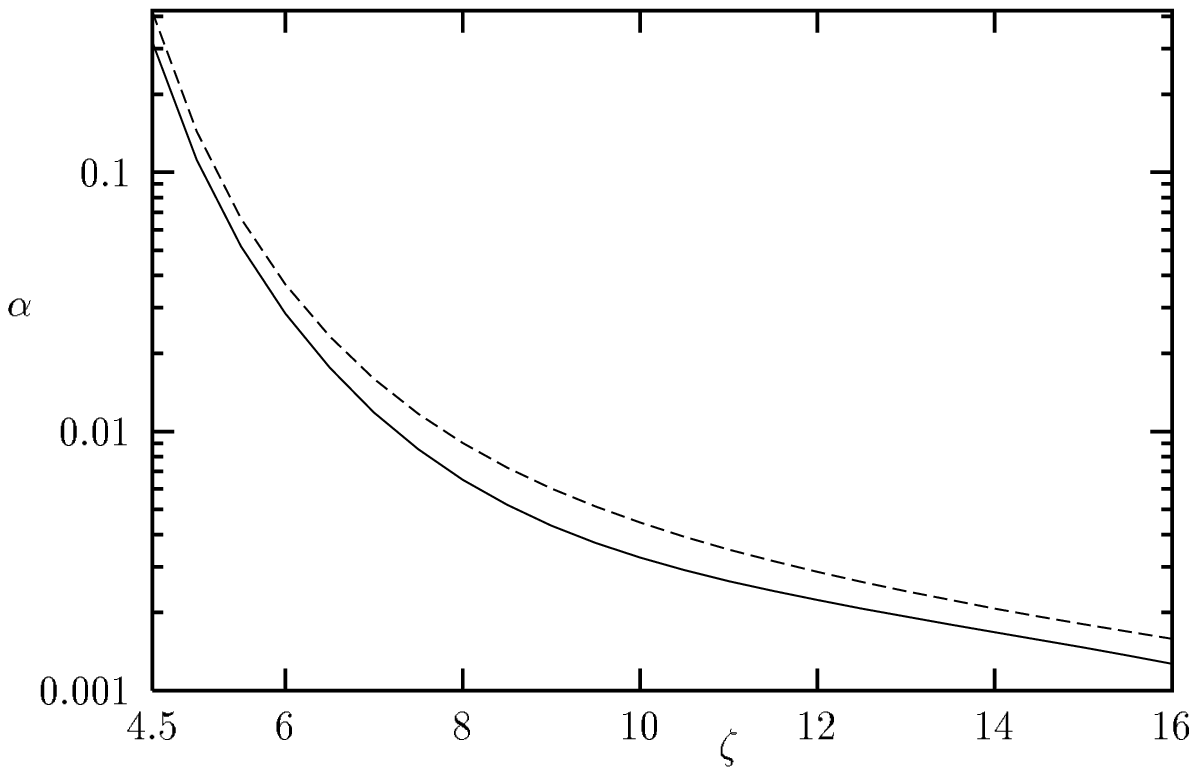}
    \vspace{-8.0cm}
\vspace{0.0cm}
\mbox{ \hspace{0.5cm} \large{\bf Figure 3: Dependence of $\alpha$ on
$\zeta$ (solid line:}}
\mbox{ \hspace{0.5cm} \large{\bf one-loop, dashed line: tree-level)}}
\end{figure}
\begin{figure}
\vspace{12cm}  
\end{figure}
\begin{figure} 
\vspace{12.0cm}
    \includegraphics{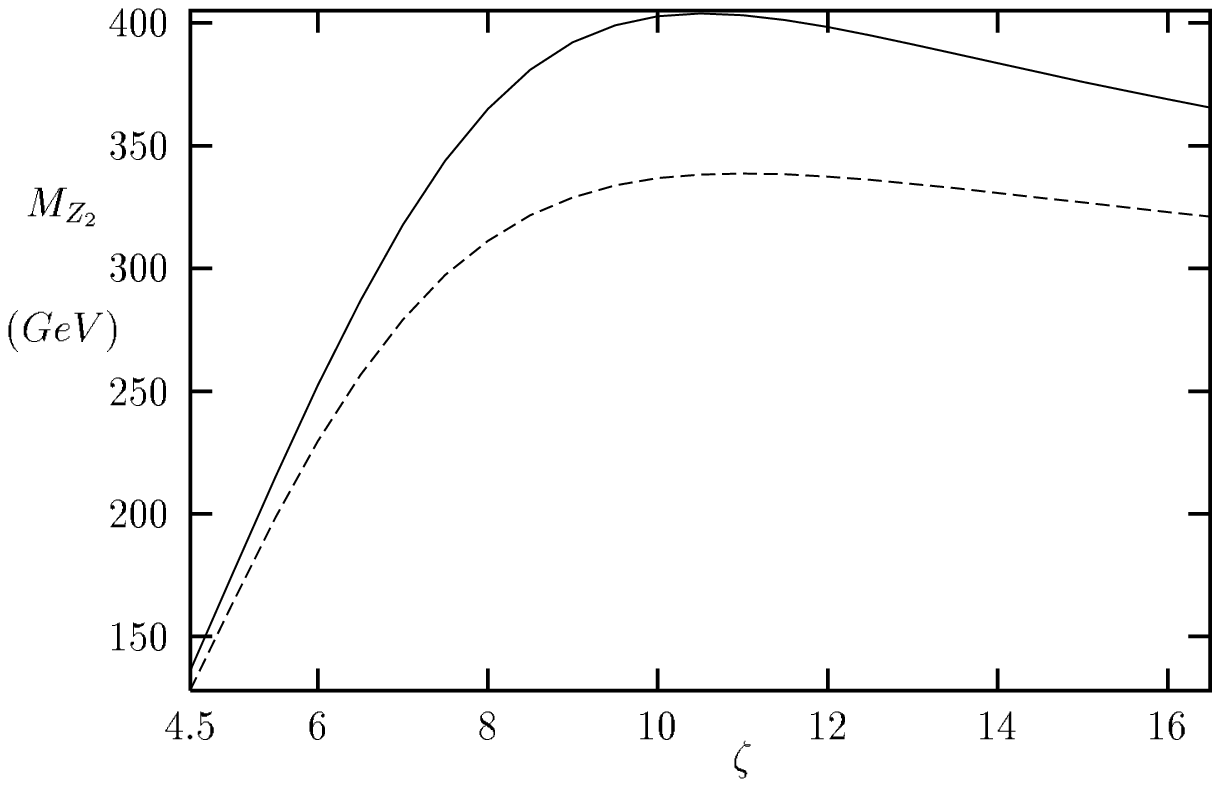}
    \vspace{-8.0cm}    
\vspace{0.0cm}
\mbox{ \hspace{0.5cm} \large{\bf Figure 4: Dependence of $M_{Z_{2}}$ on
$\zeta$ (solid line:}} 
\mbox{ \hspace{0.5cm} \large{\bf one-loop, dashed line: tree-level)}}
\end{figure}
\begin{figure}
\vspace{12cm}
\end{figure}
\begin{figure}
\vspace{12.0cm}
    \includegraphics{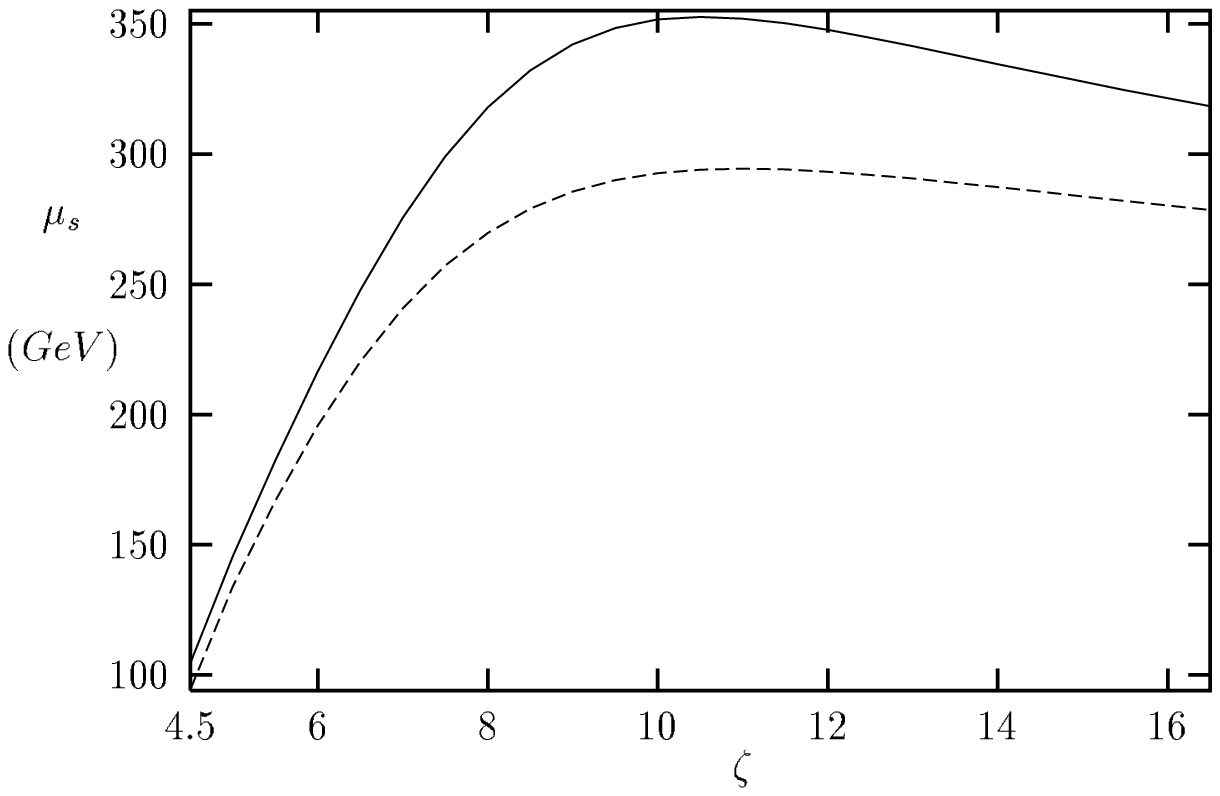}
    \vspace{-8.0cm}
\vspace{0.0cm}
\mbox{ \hspace{0.5cm} \large{\bf Figure 5: Dependence of $\mu_{s}$ on
$\zeta$ (solid line:}}
\mbox{ \hspace{0.5cm} \large{\bf one-loop, dashed line: tree-level)}}
\end{figure}
\begin{figure}
\vspace{12cm}
\end{figure}
\begin{figure}
\vspace{12.0cm}
    \includegraphics{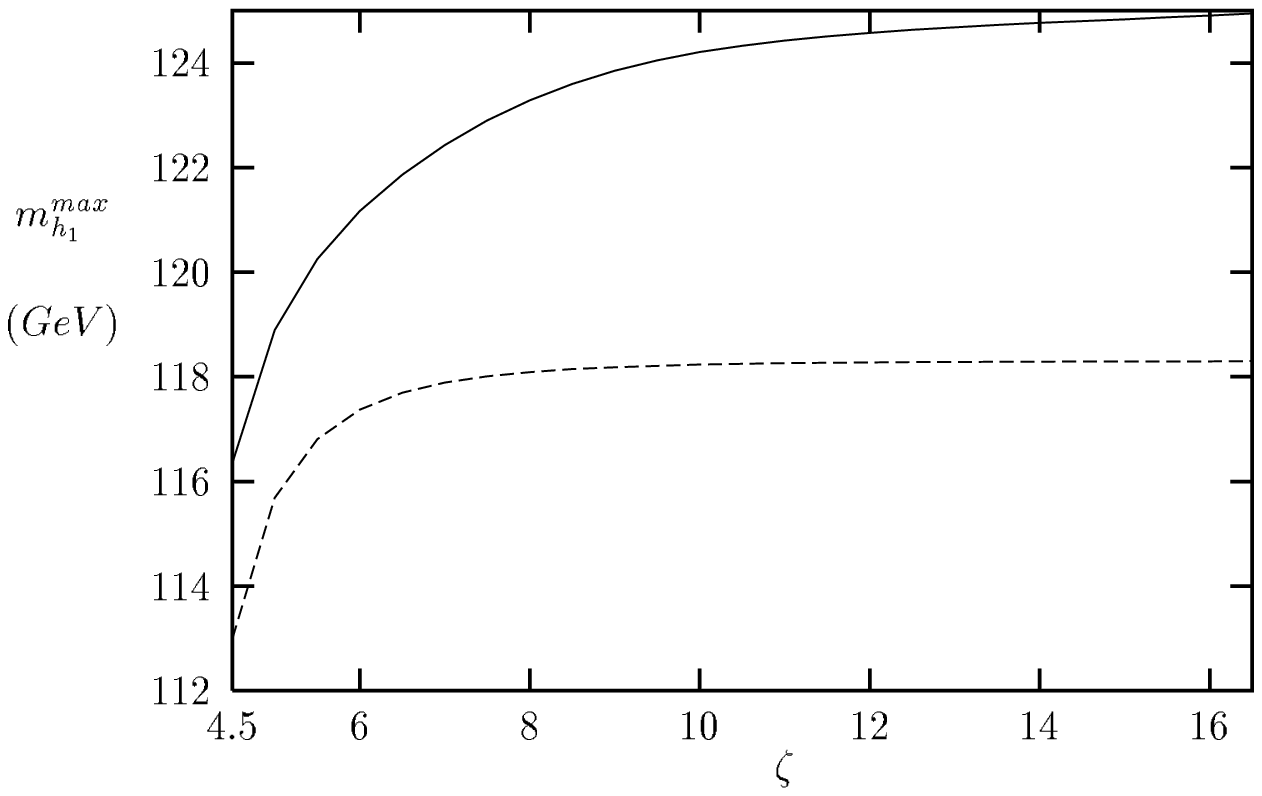}
    \vspace{-8.0cm}
\vspace{0.0cm}
\mbox{ \hspace{0.5cm} \large{\bf Figure 6: Dependence of $m_{h_{1}}^{max}$ on
$\zeta$ (solid line:}}
\mbox{ \hspace{0.5cm} \large{\bf one-loop, dashed line: tree-level)}}
\end{figure}
\end{document}